\documentstyle[twoside,fleqn,espcrc2]{article}

\def\bc{\begin{center}}
\def\ec{\end{center}}
\def\be{\begin{equation}}
\def\ee{\end{equation}}
\def\myappendix{\par
 \setcounter{section}{0}
 \setcounter{subsection}{0}
 \setcounter{equation}{0}
 \setcounter{table}{0}
 \def\appendixname{Appendix}
 \def\appesection{\setcounter{equation}{0}\section}
 \def\@thesection{\Alph{section}}
 \def\thesection{\appendixname\hskip 1.10ex\Alph{section}}
 \def\thesubsection{\@thesection.\arabic{subsection}}
 \def\theequation{\@thesection.\arabic{equation}}
 \def\thetable{\@thesection.\arabic{table}}}
\newcommand{\np}{Nucl. Phys. }
\newcommand{\pl}{Phys. Lett. }

\newcommand{\nn}{\nonumber}

\newcommand{\beq}{\begin{equation}}
\newcommand{\eeq}{\end{equation}}
\newcommand{\beqn}{\begin{eqnarray}}
\newcommand{\eeqn}{\end{eqnarray}}

\def\vdir{v\kern-7.8pt\Big{/}}
\def\pdir{p\kern-7.8pt\Big{/}}

\title{Non-leptonic weak decays and final state interactions in lattice
QCD\thanks{Work done
in collaboration with M. Ciuchini, E. Franco and G. Martinelli}
\thanks{Talk given at the "QCD 96" Conference, Montpellier (France) July
4-12th 1996.}}
\author{L. Silvestrini \address{Dipartimento di Fisica, Universit\`a di Roma
	{\it ``Tor Vergata"} and INFN, Sezione di Roma II, \\
        via della Ricerca Scientifica, 1 , I-00133 Roma, Italy} }

\begin{document}
\pagestyle{empty}
\begin{abstract}
We show that, under a reasonable ``smoothness" hypothesis, it is possible to
extract informations on the amplitude and phase of two-body non-leptonic
weak decay matrix elements from the study of Euclidean correlation functions
in lattice QCD.
\end{abstract}

\maketitle

\section{INTRODUCTION} \label{intro}
Exclusive non-leptonic weak decays are a fundamental ingredient in the
determination of the Cabibbo-Kobayashi-Maskawa (CKM) matrix elements and in
the study of CP violation in the K-, D- and B-meson decays. However, the
study of these non-leptonic decays involves non-perturbative aspects of the
strong interactions, a challenging task for  our still incomplete knowledge
of non-perturbative dynamics.
Unfortunately,
a theoretical description of exclusive decays based on the fundamental theory
is not  possible yet. Over the years, several methods have been introduced
 to estimate
the relevant matrix elements: vacuum saturation, bag models,
quark models,  QCD sum rules,
$1/N_c$ expansion, chiral Lagrangians, factorization, etc.
However, none of these approaches is actually based on first principles.
\par The lattice approach has been used
to obtain results based on first principles
  for a wide set of relevant physical quantities such as
the hadron spectrum, the meson decay constants,
the form factors entering in semileptonic and radiative decays,
the kaon $B$-parameter $B_K$, etc. However, any computation of
exclusive non-leptonic decays performed in the Euclidean space-time suffers
from severe limitations, as shown by Blok and Shifman \cite{bs}
in the context of QCD sum rules, and by Maiani and Testa \cite{mt} in the
framework of lattice QCD. In fact, the activity
in this field \cite{bss,ab1}
has completely stopped after the publication of ref.~\cite{mt}.
\par In this paper we show that, under quite reasonable  physical hypotheses,
 it is possible, at least in principle, to extract predictions
for the relevant matrix elements in numerical simulations
of lattice QCD, in spite of the difficulties due to the Maiani-Testa No-Go
Theorem (MTNGT).
\par The MTNGT states essentially the following:
\begin{itemize} \item In the calculation
of a two- (many-) body decay amplitude performed
in the Euclidean space-time, which is
the only possibility in Monte Carlo simulations,
 there is no distinction between {\it in-} and
{\it out-}states. As a consequence, the matrix elements that one is able
to extract are real numbers resulting from the  average of the two cases.
This jeopardizes the possibility
of  any realistic prediction for the matrix elements. For example, we know from
the measured $A_{1/2}$ and $A_{3/2}$ amplitudes in $D \to K\pi$ decays
that there is a phase difference of about $80^\circ$.
\item Matrix elements are extracted on the lattice by studying the
time behaviour of appropriate correlation functions at  large time
distances. Maiani and Testa showed  that what can be really isolated
in this limit are the off-shell form factors corresponding to
the final particles at rest.
For kaon decays, we can use the chiral theory to extrapolate the
form factor to the physical point.
This is certainly not the case for $D$- and $B$-meson
decays. In the latter case
it is not possible to obtain a realistic prediction for the matrix element.
\end{itemize}
\par We will show that both these difficulties
can be overcome under a ``smoothness'' hypothesis,
and that under this hypothesis it is possible, at least in principle,
to extract the physical matrix elements, including the phase due
to the strong-interaction rescattering of the final states.
In the interesting case in which
Final State Interaction (FSI) is dominated by the exchange of a resonance in
the $s$-channel, it is possible to calculate the parameters of the resonance,
as
explained in detail in ref.~\cite{noi}.
\par Unfortunately, at present
 we do not know if the method we are proposing can be
succesfully applied in numerical simulations on currently available
lattices.

\section{CORRELATION FUNCTIONS IN THE EUCLIDEAN SPACE-TIME}
Following ref.~\cite{mt}, we first examine the Euclidean three-point function
$G_{\vec q}(t_1,t_2)$:
\beqn
G_{\vec q}(t_1,t_2)&\equiv &
\langle 0 \vert  T \left[ \Pi_{\vec q}(t_1) \Pi_{-\vec q}(t_2) H(0)
\right]  \vert 0 \rangle\nn \\
&=&\langle 0 \vert \Pi_{\vec q}(t_1) \Pi_{-\vec q}(t_2) H(0)
\vert 0 \rangle
\label{eq:3pf}
\eeqn
when $t_1>t_2>0$.
In eq.~(\ref{eq:3pf}),  $\Pi_{\vec q}(t)$ is an interpolating field
of the final-state particle (denoted as  ``pion"
in the following) with a fixed spatial momentum
\be \Pi_{\vec q}(t) = \int \,  d^3x \,  e^{- i\vec q \cdot \vec x}
\Pi (\vec x , t)\, ; \ee
$H(0)=H(\vec x=0, t=0)$ is any local operator that couples to the
two pions in the final state; $ T \left[ \dots \right]$ represents
the $T$-product of the fields and   the vacuum
expectation value corresponds, in a numerical simulation, to the
average over the gauge field configurations. \par
When $t_1 \to \infty$
\beqn G_{\vec q}(t_1,t_2)&\to& \sum_n \langle 0\vert  \Pi_{\vec q}(t_1)
\vert n \rangle \langle n \vert \Pi_{-\vec q}(t_2) H(0) \vert 0 \rangle \nn \\
&\sim& \frac{\sqrt{Z_\Pi}}{2 E_{\vec q}} e^{-E_{\vec q}t_1} G_3(t_2)
\, ,
\eeqn
where
\be E_{\vec q}=\sqrt{M_\pi^2+{\vec q}^2}\,,\quad
\sqrt{Z_\Pi} =\langle 0 \vert \Pi(0) \vert \vec q \rangle \ee
and
\beq G_3(t)=\langle \vec q\, \vert \Pi_{-\vec q}(t) H(0)\vert 0 \rangle \eeq
for $t>0$.
\par Inserting a complete set of {\it out}-states $n$, we can write
\beqn
&\,&\langle \vec q \, \vert \Pi_{-\vec q}(t) H(0)\vert 0 \rangle
=  \sum_n (2 \pi)^3
\delta^{(3)}(P_n) \times \nn \\
&\,&\langle \vec q \, \vert\Pi(0)
\vert n \rangle \langle n \vert H(0) \vert 0 \rangle
e^{-(E_n-E_{\vec q}) t} = \nn \\
&\,&\frac{\sqrt{Z_\Pi}}{2E_{\vec q}} e^{-E_{\vec q} t}
\,_{out}\langle \vec q, -\vec q \vert H(0) \vert 0 \rangle +
\sum_n (2 \pi)^3 \times \nn \\
&\,& \delta^{(3)}(P_n) \langle \vec q\, \vert\Pi(0) \vert n  \rangle^{c}
\langle n
 \vert H(0) \vert 0 \rangle
e^{-(E_n-E_{\vec q}) t}
\label{starting}
\eeqn
where  the  term
proportional to $_{out}\langle \vec q,- \vec q \vert H(0) \vert 0\rangle$
on the r.h.s. is the disconnected
contribution.
We now define the connected matrix element of the pion field as follows:
\beq
\langle \vec q \,\vert\Pi(0) \vert n  \rangle^{c}=
\frac{2\sqrt{Z_\Pi}[{\cal M}(\vec q,- \vec q;n)]^{*}}{
(E_n+2 E_{\vec q})
(-E_n+2 E_{\vec q}-i \epsilon)}
\label{defi}
\eeq
The LSZ reduction formula shows that, when the four momentum of the pion field
$\Pi(0)$ goes on the mass-shell ($E_n \to 2 E_{\vec q}$), ${\cal M}$ reduces to
the invariant scattering amplitude of the process $\pi(\vec q) + \pi(-\vec q)
\to n$, cf. eq.~(15) of ref.~\cite{mt}:
\beqn
&\,&(2 \pi)^4 \delta^{(3)}(P_n)\delta(E_n-2 E_{\vec q})[i{\cal M}(\vec q,- \vec
q;n
)]^{*}=\nn \\
&\,&_{in}\langle\vec q,- \vec q\vert n\rangle_{out}-_{out}\langle\vec q,-
\vec q\vert n\rangle_{out}\, .
\eeqn
In
general one could write
\beqn
  &\,&\langle n \vert \Pi(0)
 \vert \vec q \rangle =\,\,\sqrt{Z_\Pi} {\cal M}(\vec q,- \vec q;n)
  {\cal F}\left(\frac{E_n}{2E_{\vec q}}\right)\nn \\
&\,&\times \left[ \frac{2}
{(E+2 E_{\vec q}-i \epsilon)
(-E+2 E_{\vec q}-i \epsilon)}\right],
\label{defig}
\eeqn
with the condition that the modulating factor ${\cal F}$
satisfies
${\cal F}(1)=1$ for the on-shell pion ($E_n/2E_{\vec q}=1$).
The factor in square brackets is (up to a factor
$1/2 E_{\vec q}$) the propagator of two non-interacting
pions \cite{cm}.

Inserting the definition (\ref{defi}) into eq.~(\ref{starting}), and using the
identity
\[ \frac{1}{E-2E_{\vec q}-i\epsilon}= {\cal P}\left[
\frac{1}{E-2E_{\vec q}}\right]+i\pi\delta(E-2E_{\vec q})\,
,  \]
 we obtain
\beqn
&\,&G_3(t)=\frac{\sqrt{Z_\Pi}}{2 E_{\vec q}} e^{E_{\vec q}t}
\Biggl[P_{\vec q}(t) + \frac{1}{2} \times \\
&\times&\biggl(\,_{out}\langle \vec q, -\vec q
\vert H(0) \vert 0 \rangle +
\, _{in}\langle \vec q, -\vec q
\vert H(0) \vert 0 \rangle \biggr) \Biggr] , \nn
\label{result1}
\eeqn
with
\beqn
P_{\vec q}(t) &=&(4E_{\vec q}) {\cal P}\sum_n e^{-(E_n - 2 E_{\vec q})t} (2
\pi)^3 \delta^{(3)}(P_n) \nn \\
&\times&
\frac{ {\cal M}^{*}(\vec q,- \vec q;n)\langle n  \vert H(0) \vert 0 \rangle }{
4E_{\vec q}^2-E_n^2},
\eeqn
which corresponds to eq.~(21) of ref.~\cite{mt}.
$P_{\vec q}(t)$ is a sum over off-shell amplitudes. In the limit $t \to
+\infty$, it is dominated by intermediate states with energy $E_n < 2 E_{\vec
q}$, which correspond to two-pion states with momenta $\vec k$ and $-\vec k$,
with $\vert \vec k \vert < \vert \vec q \vert$.
We therefore obtain, for $t \gg 0$,
\beqn
P_{\vec q}(t) &=& (4E_{\vec q}) {\cal P} \int \frac{d^3 k_1}{(2 \pi)^3 2
E_{\vec{k}_1}} \frac{d^3 k_2}{(2 \pi)^3 2 E_{\vec{k}_2}}  \nn \\
&\times&(2 \pi)^3 \delta^{(3)}(\vec{k}_1 + \vec{k}_2) e^{-(E_{\vec{k}_1} +
E_{\vec{k}_2} - 2 E_{\vec q})t}  \nn \\
&\times& \frac{ {\cal M}^{*}(\vec q,- \vec q;\vec k_1,\vec k_2)\langle \vec
k_1,\vec k_2  \vert H(0) \vert 0 \rangle }{
4E_{\vec q}^2-(E_{\vec{k}_1}+E_{\vec{k}_2})^2} = \nn \\
&=& (4E_{\vec q}){\cal P} \int_{2 M_{\pi}}^\infty \frac{dE}{2 \pi}
\frac{e^{-(E - 2 E_{\vec q})t}}{4E_{\vec q}^2-E^2} \nn \\
&\times& (2 \pi)^4 \int \frac{d^3 k_1}{(2 \pi)^3 2
E_{\vec{k}_1}} \frac{d^3 k_2}{(2 \pi)^3 2 E_{\vec{k}_2}} \nn \\
&\times&  \delta^{(3)}(\vec{k}_1 + \vec{k}_2)
\delta(E-E_{\vec{k}_1}-E_{\vec{k}_2}) \nn \\
&\times&  {\cal M}^{*}(\vec q,- \vec q;\vec k_1,\vec k_2)\langle \vec k_1,\vec
k_2  \vert H(0) \vert 0 \rangle  = \nn \\
&=& (4E_{\vec q}){\cal P} \int_{2 M_{\pi}}^\infty \frac{dE}{2 \pi}
\frac{e^{-(E - 2 E_{\vec q})t}}{4E_{\vec q}^2-E^2} \nn \\
&\times& \frac{1}{16 \pi} \sqrt{\frac{E^2-4 M_{\pi}^2}{E^2}}
\biggl[{\cal M}^{*}(\vec q,- \vec q;\vec k,-\vec k)\nn \\
&\times&\langle \vec k,-\vec k  \vert H(0) \vert 0 \rangle\biggr]_{\vert k
\vert=\sqrt{E^2-4 M_{\pi}^2}/2}\, .
\label{pq1}
\eeqn
In the last equality, we have only considered $s$-wave two-pion scattering, and
therefore integrated over the angles in the phase space\footnote{In general,
one should
also include partial amplitudes with non-zero angular momentum, and retain the
angular dependence. The generalization of the above formulae to the case of
total momentum and angular momentum different from zero is straightforward.}.

Watson theorem states that, in absence of CP violation,
\beq \,_{out}\langle \vec k,-\vec k \vert H(0)
 \vert 0 \rangle = e^{i\delta(s)} A(s)\,
\label{out}\eeq
and
\beq \,_{in}\langle \vec k,-\vec k \vert H(0)
 \vert 0 \rangle = e^{-i\delta(s)} A(s)\,
\label{in}\eeq
with $A(s)$ real. The phase $\delta$ is related to the two-pion scattering
amplitude by
\beq _{out}\langle \vec q,-\vec q \vert \vec q,-\vec q
\rangle_{in}=  e^{2 i \delta(s)}\, .
\label{phase} \eeq
In the case of {\em{on-shell}} $s$-wave $\pi - \pi$ scattering, the invariant
matrix element is given in terms of the phase by
\beq
{\cal M}(s)=\frac{(16 \pi) \sqrt{s}}{\sqrt{s-4 M_{\pi}^2}} e^{i \delta(s)}\sin
\delta(s) \, .
\eeq
We now make use of our ``smoothness" hypothesis to write the off-shell matrix
element as
\beq
{\cal M}(\vec q,- \vec q;\vec k,-\vec k)=
\frac{(16 \pi) \sqrt{s}}{\sqrt{s-4 M_{\pi}^2}} e^{i \delta(s)}\sin \delta(s) \,
,
\label{mofs}
\eeq
with $s=4 E_{\vec k}^2$. Inserting the definitions (\ref{out}) and (\ref{mofs})
into eqs.~(\ref{result1}) and (\ref{pq1}), we obtain
\[
\frac{P_{\vec q}(t)}{4E_{\vec q}}= {\cal P}\int_{2 M_{\pi}}^\infty \frac{dE}{
\pi}
\frac{e^{-(E - 2 E_{\vec q})t}}{4E_{\vec q}^2-E^2} A(E^2) \sin \delta(E^2)
\]
and
\beqn
G_3(t)=\frac{\sqrt{Z_\Pi}}{2 E_{\vec q}} e^{E_{\vec q}t}
\Biggl[A(s_q) \cos \delta(s_q) + (4E_{\vec q}) \times \nn \\
{\cal P}\int_{2 M_{\pi}}^\infty \frac{dE}{ \pi}
\frac{e^{-(E - 2 E_{\vec q})t}}{4E_{\vec q}^2-E^2} A(s) \sin \delta(s)
\Biggr]
\label{g3res}
\eeqn
with $s_q=4 E_{\vec q}^2$ and $s = E^2$.
\section{MESON DECAYS ON THE LATTICE}
To fix our ideas, let us consider D decays. In this case, the starting point is
 the four-point correlation function
\beqn
&\,&G(t_1,t_2,t_D,\vec q,-\vec q,\vec p_D=0)= \nn \\
&=&\langle \Pi_{\vec q}(t_1)
 \Pi_{-\vec q}(t_2) {\cal H_W}(0)
 D^\dagger_{\vec p_D=0}(t_D) \rangle
\label{Wdecay}
\eeqn
in the limit $t_D \to - \infty$, $t_1 \to + \infty$. Here  ${\cal H}_W$ is the
weak Hamiltonian. In this case eq.~(\ref{g3res}) becomes
\beqn
G(t_1,t_2,t_D,\vec q,-\vec q,\vec p_D=0) =
\frac{Z_\Pi}{4E^2_{\vec q}} \frac{\sqrt{Z_D}}{2M_{D}}
\times \nn \\
e^{-E_{\vec q}(t_1+t_2)-M_{D}\vert t_D\vert}
\Biggl\{ A^W(s_q) \cos\delta(s_q) + (4 E_{\vec q})
 \nn \\
\times{\cal P} \biggl[
\int_{2 M_\pi}^{\infty}
\frac{dE}{\pi} \frac{e^{-(E-2E_{\vec q})t_2} }{4E_{\vec q}^2-E^2}
A^W(s)\sin\delta(s) \biggr]\Biggr\}.\nn
\label{ddelta}
\eeqn
\par For numerical applications, it is convenient
to consider the amputated correlation function given by
the ratio
\[
R^{{\cal H}_W}(t_2,\vec q) =
\frac{G(t_1,t_2,t_D,\vec q,-\vec q,\vec p_D=0)}{S_\Pi(t_1,E_{\vec q})
S_\Pi(t_2,E_{\vec q})S_D(t_D, M_D)}\, ,
\]
where \beq S_\Pi(t_1,E_{\vec q})=\frac{\sqrt{Z_\Pi}}{2E_{\vec q}}
e^{-E_{\vec q}t_1}\, ,
\label{ratio}
\eeq
and similarly for the other meson propagators;
\beqn
R^{{\cal H}_W}(t_2,\vec q) =
\Biggl\{A^W(s_q) \cos\delta(s_q) +(4 E_{\vec q}) &\times& \nn \\
{\cal P} \biggl[
\int_{2 M_\pi}^{+\infty}
\frac{dE}{\pi}\frac{e^{-(E-2E_{\vec q})t_2} }{4E_{\vec q}^2-E^2}
A^W(s)\sin\delta(s) \biggr]\Biggr\}.&\,&
\label{example}
\eeqn

On a lattice of volume $L^3 \times T$, eq.~(\ref{example}) becomes
\beqn &\,& R^{{\cal H}_W}(t_2,\vec q) =
A( s_q) \cos\delta(s_q) + \left(\frac{4 E_{\vec q}}{\pi}\right) \times  \nn
\\
&\,&{\sum_{E_i}^{}}^\prime\left[ \Delta E_i
 \frac{e^{-(E_i- 2 E_{\vec q})t_2}}{
4 E_{\vec q}^2-E_i^2}
A(s)\sin\delta(s) \right] ,
\label{examplel}
\eeqn
where all the quantities are given in units of the lattice spacing,
 \beqn E_i = \sqrt{s}=
2 E_{\vec k} \quad \mbox{with} \quad
\vec k\equiv \frac{2 \pi}{L}(n_x,n_y,n_z) \,  ;\eeqn
$s_q$ has been defined before; $n_{x,y,z}=0,1,\dots,L-1$ and
 $\sum_{E_i}^\prime$ denotes the sum over all the
values of the energy corresponding to  the momenta $\vec k$ allowed by
the discretization of the space-time on a finite volume, excluding those
corresponding to
$E_i= 2 E_{\vec q}$. Different
combinations of momenta corresponding to the same energy should be included
only
once in the sum appearing in eq.~(\ref{examplel}).
$\Delta E_i= E_{i+1}-E_i$ is the difference of the nearest successive
allowed values of $E_i$ ($E_0 = 2 M_\pi$, $E_1=
2\sqrt{M_\pi^2+ (2 \pi/La)^2}$, $E_2=2\sqrt{M_\pi^2+2 (2 \pi/La)^2}$,
etc.). One can show that the expression in eq.~(\ref{examplel}) tends to
the corresponding continuum one in eq.~(\ref{example}) as $L \to \infty$.
However, for presently available lattices, the allowed range of the momenta
and of $s_q$ is limited, and therefore eq.~(\ref{examplel}) might not be a
good approximation of eq.~(\ref{example}). One might try to improve the
approximation by using $\vec p_D \neq 0$ to increase the allowed values of the
momenta, and by correcting for the difference between the discrete phase
space and the continuum one.

We now explain the strategy to extract the matrix element of ${\cal H}_W$
from $R^{{\cal H}_W}(t_2,\vec q)$. To fix our ideas, we consider the case $
\vec p_D =0$. The procedure is the following:
\begin{enumerate}
\item Compute $R^{{\cal H}_W}(t_2,\vec q=0)$. The phase vanishes at
threshold, and therefore we have
\beq
R^{{\cal H}_W}(t_2,\vec q=0)=A^W(4 M_\pi^2)\, ,
\eeq
and we can extract $A^W(4 M_\pi^2)$.
\item Compute $R^{{\cal H}_W}(t_2,\vec q_1)$ for the first allowed non-zero
value of the momentum, $\vert q_1 \vert = (2 \pi)/L$. The only value of $k$
that might contribute in the sum in eq.~(\ref{examplel}), apart from
exponentially suppressed terms, is $k=0$, but at threshold the phase
vanishes and therefore there is no contribution from the sum over $k$. Thus
we have
\beq
R^{{\cal H}_W}(t_2,\vec q_1)=A^W(4 E_1^2)\cos \delta(4 E_1^2)\, ,
\eeq
with $E_1=\sqrt{M_\pi^2+(2 \pi)^2/(La)^2}$.
\item Compute~$R^{{\cal H}_W}(t_2,\vec q_2)$,~with $\vert q_2 \vert =
\sqrt{2}(2\pi)/L$. In this case, the term in the sum over $k$ corresponding
to $\vert k \vert = (2 \pi)/L$ gives an exponentially increasing
contribution, and we get
\beqn
R^{{\cal H}_W}(t_2,\vec q_2)=A^W(4 E_2^2)\cos \delta(4 E_2^2)
+\frac{E_2}{\pi}\times &\,& \nn \\
(E_2-E_1)\frac{e^{2 (E_2 - E_1)t_2}}{E_2^2 - E_1^2}
A^W(4 E_1^2)\sin \delta(4 E_1^2),&\,& \nn
\eeqn
where $E_2=\sqrt{M_\pi^2+2(2 \pi)^2/(La)^2}$.
\end{enumerate}
It is straightforward to derive the expression of $R^{{\cal H}_W}$ for the
next steps. In this way, we can extract
\beq
\tan\delta(s)=\frac{A_s}{A_c} \quad
\mbox{and}\, \,\,\, \vert A(s)\vert=\sqrt{A_c^2+A_s^2}
\eeq
as a function of the centre-of-mass energy,
where $A_c=A(s) \cos \delta(s)$ and $A_s=A(s)
 \sin \delta(s)$.

An interesting case, which has been discussed in detail in ref.~\cite{noi},
is the one of FSI's dominated by the exchange of a resonance $\sigma$ in the
$s$-channel\footnote{In ref.~\cite{noi}, the result corresponding to
eq.~(\ref{g3res}) of the present work has been derived in the case of FSI's
dominated by the exchange of a resonance. However, as we have shown above,
the result (\ref{g3res}) holds under a more general ``smoothness" hypothesis.}.
In this case, we can express the matrix elements of $H$ and
$\Pi$ in terms of the mass of the resonance $M_\sigma$, its width
$\Gamma=X/M_\sigma$ and its couplings $g$ and $V$, to $H$ and to the two-pion
state respectively:
\[
_{out}\langle \vec q,-\vec q \vert H(0) \vert 0 \rangle
=\frac{g(s_q)}{
M^2_\sigma-s_q-i X(s_q)}\vert_{s_q=p_H^2}\, ,
\]
\[
\langle \vec q \,\vert\Pi(0) \vert\vec k,-\vec k  \rangle_{out}
=
\biggl[ \frac{\vert V(s) \vert^2}{M^2_\sigma-s-i X(s)} \times
\]
\beq
\times\frac{2\sqrt{Z_\Pi}}{(E+2 E_{\vec q}-i \epsilon)
(-E+2 E_{\vec q}-i \epsilon)}
\biggr]^*\,.
\label{defis}
\eeq
The phase shift and the amplitude are given in terms of the parameters of
the resonance by the following relations:
\beq
\frac{g(s)}{M^2_\sigma-s-i X(s)}=A(s)e^{i\delta(s)}
 \label{due}
\eeq
and
\beqn  \cos \delta(s)&=&\frac{M^2_\sigma-s}{\sqrt{(M^2_\sigma-s)^2+X(s)^2}}
\\
 \sin \delta(s)&=&\frac{X(s)}{\sqrt{(M^2_\sigma-s)^2+X(s)^2}} \, .\eeqn
By studying the three-point $H-\Pi-\Pi$ correlator $G_{\vec q}(t_1,t_2)$, we
can extract the parameters of the resonance and exploit this information
when analyzing the four-point function.

\section{CONCLUSIONS}
We have shown that, in spite of the MTNGT, it is in principle
possible to extract the
amplitude and phase of two-body non-leptonic decay matrix elements, under a
reasonable ``smoothness" hypothesis. We have sketched the strategy to
extract the relevant information from Euclidean correlators computed
numerically in lattice QCD. In the case of FSI's dominated by a resonance, it
is possible to extract the parameters of the resonance from the study of a
three-point correlation function. A feasibility study of the method we
propose is
currently under way on the APE machine.

\section*{ACKNOWLEDGEMENTS}
The results presented here have been obtained in a most enjoyable
collaboration with M. Ciuchini, E. Franco and G. Martinelli, to whom I am
much indebted. I would also like to thank G. Martinelli for his help in
preparing my transparencies, and E. Franco for reading the manuscript.
I warmly thank S. Narison for organizing such a stimulating conference.

\end{document}